\documentclass{edbk}
\usepackage{epsfig}
\let\footnote\savefootnote
\let\footnoterule\savefootnoterule 
\normallatexbib

\begin{document}
\articletitle{Fermi surface, pseudogaps and \\ 
dynamical stripes in
L\lowercase{a}$_{2-\lowercase{x}}$S\lowercase{r}$_{\lowercase{x}}$C\lowercase{u}O$_4$}

\author{A. Fujimori$^1$, A. Ino$^2$, T. Yoshida$^1$, T.
Mizokawa$^1$}

\affil{$^1$Department of Physics and Department of
Complexity Science and Engineering, University of Tokyo,
Bunkyo-ku, Tokyo 113-0033, Japan\\
$^2$Japan Atomic Energy Research Institute, SPring-8,
Mikazuki, Sayo, Hyogo 679-5198, Japan}

\author{Z.-X. Shen, C. Kim}

\affil{Department of Applied Physics and Stanford
Synchrotron Radiation Laboratory, Stanford University,
Stanford, CA94305, U.S.A.}

\author{T. Kakeshita, H. Eisaki, S. Uchida}

\affil{Department of Advanced Materials Science, University
of Tokyo, Bunkyo-ku, Tokyo 113-0033, Japan}

\begin{abstract}
Doping dependence of the electronic structure of
La$_{2-x}$Sr$_x$CuO$_4$ (LSCO) has been systematically
studied in a series of photoemission measurements.  The
unusual spectral features in the underdoped regime are
attributed to the formation of dynamical stripes and the
opening of large and small pseudogaps.

\end{abstract}

\section{Introduction}

The most remarkable feature in the high-$T_c$ cuprates is their characteristic
phase diagram as a function of hole doping or band filling, which covers 
from the antiferromagnetic insulating phase near the undoped limit 
to the normal Fermi-liquid phase in the overdoped regime 
with the intervening superconducting phase. 
Furthermore, in the underdoped  
superconducting phase, ``non-Fermi-liquid" properties such as 
pseudogap behaviors are observed. 

In order to systematically understand 
the origin of the phase diagram and the nature of each phase, 
La$_{2-x}$Sr$_x$CuO$_4$ (LSCO) is a unique 
system in that it covers the whole range of the phase diagram in a single
system. In addition, it 
has the simplest crystal structure with single CuO$_2$ layers and 
its hole concentration is rather accurately determined by 
the Sr concentration $x$ (plus small oxygen non-stoichiometry). 
On the other hand, LSCO is complicated in that 
it undergoes a structural distortion from the high-temperature
tetragonal (HTT) phase to the low-temperature
orthorhombic (LTO) phase in the superconducting compositions 
and even has an 
inherent instability towards the low-temperature tetragonal 
(LTT) phase. The latter phase is realized in 
La$_{2-x-y}$Nd$_y$Sr$_x$CuO$_4$ (LNSCO), accompanied by the 
ordering of charge and spins in a stripe form, especially 
around $x\sim1/8$ \cite{Tranquada}. 
Transport measurements of LNSCO have shown that 
the static stripes are indeed one-dimensional metals \cite{Uchida_stripe}. 
Recently, 
LNSCO with $y=0.4$ and $x=0.12$ was studied by angle-resolved photoemission 
spectroscopy (ARPES) by Zhou {\it et al.} \cite{Zhou}. They observed
Fermi surface features characteristic of 
a (quarter-filled) one-dimensional metal, 
namely, confinement of spectral weight within 
$|k_x|, |k_y|<\pi/4$. In LSCO, 
the stripes are not static but are thought to 
remain dynamical fluctuations, as reflected on the
incommensurate inelastic neutron peaks \cite{Yamada}.

In this article, we summarize the results of our photoemission studies on 
LSCO, focusing on the systematic evolution of the electronic structure as a 
function of hole doping including the formation of the pseudogaps. 
Some unusual spectral features characteristic of LSCO 
are indeed attributed to the dynamical stripes. 

\section{Band dispersion and Fermi surface}

\begin{figure}[htb]
\center{\epsfig{file=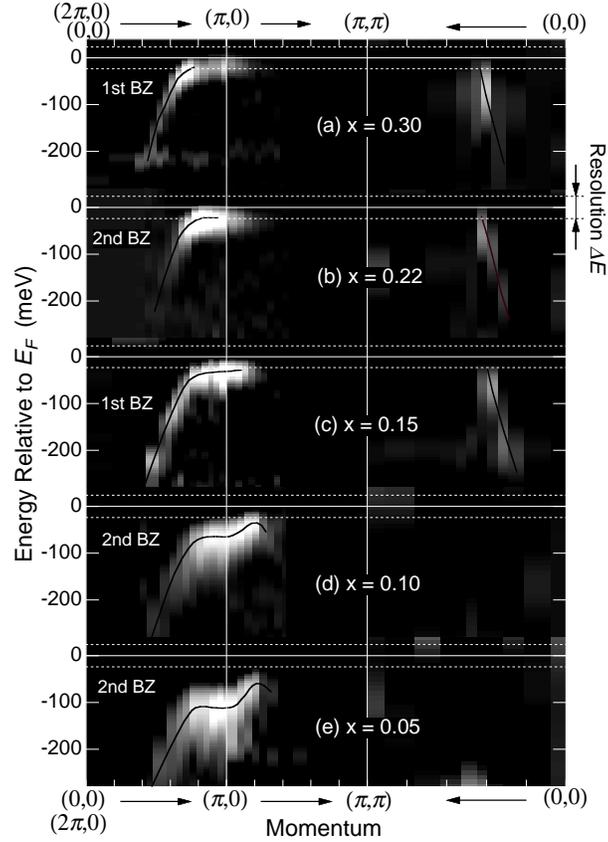,height=112mm}}
\caption{Band dispersions in LSCO \protect\cite{Ino_ARPES2}.}
\label{fig0}
\end{figure}

\begin{figure}[htb]
\center{\epsfig{file=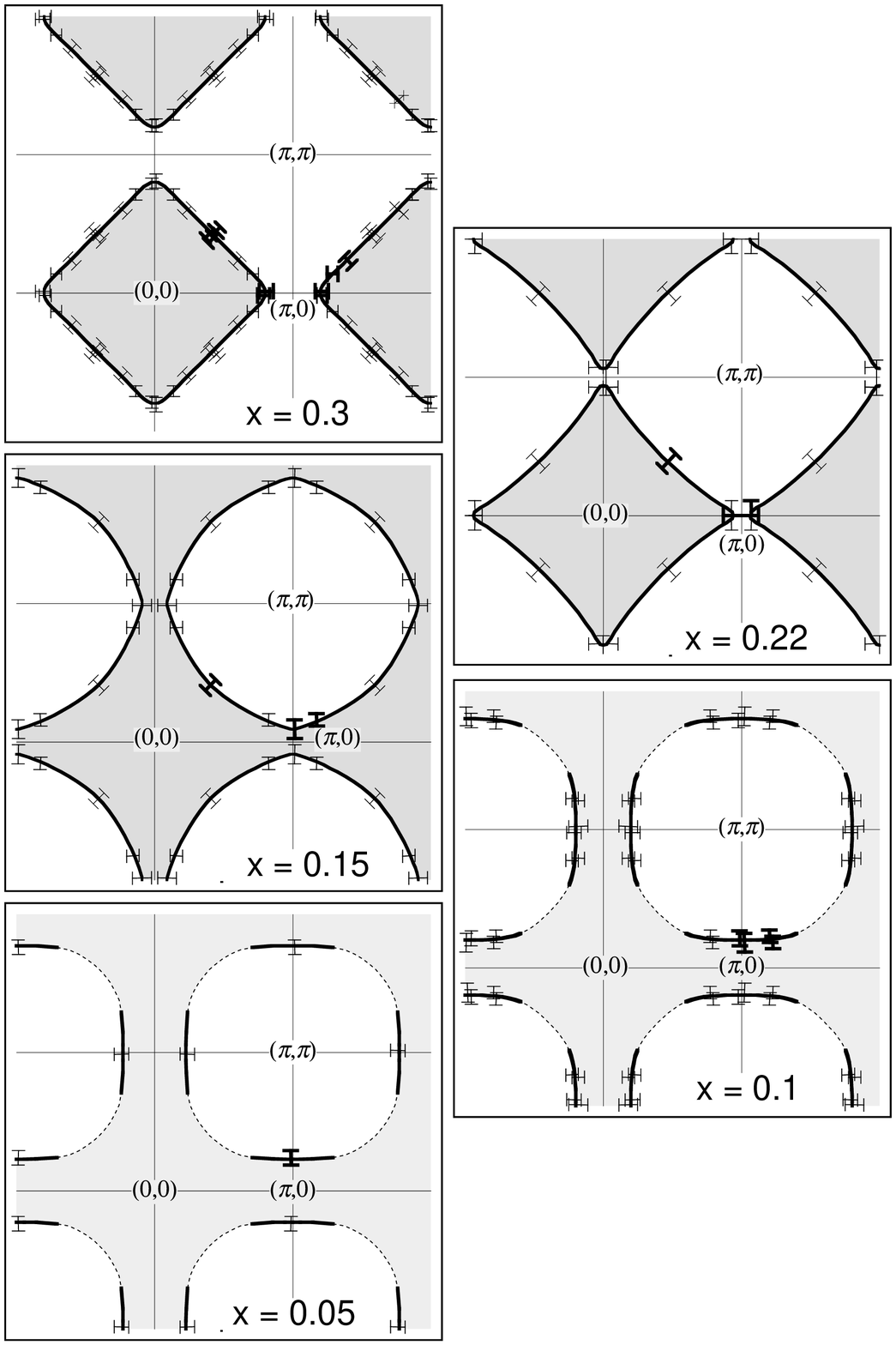,height=90mm}}
\caption{Fermi surface or minimum gap locus of LSCO determined by 
band dispersions in the ARPES spectra \protect\cite{Ino_ARPES2}.}
\label{fig1}
\end{figure}

The experimental band structure of optimally doped LSCO
$(x\simeq0.15)$ studied by ARPES \cite{Ino_ARPES,Ino_ARPES2}
is similar to that of Bi$_2$Sr$_2$CaCu$_2$O$_8$ (BSCCO),
which have been extensively studied by ARPES, as shown in
Fig.~\ref{fig0}.  The band is flat around ${\bf k}=(\pi,0)$
[especially along the $(\pi,0)$ direction] while the band
crossing the Fermi level ($E_F$) around $(\pi/2,\pi/2)$ is
strongly dispersive.  The ``flat band" rises with $x$,
crosses $E_F$ at $x\sim0.2$ and goes above $E_F$ for $x >
0.2$ \cite{Ino_ARPES2}.  With decreasing $x$, the flat band
is lowered, and around $x\sim0.1$ it becomes as low as
$\sim0.1$ eV below $E_F$ \cite{Ino_SIT}.

As the position of the flat band relative to $E_F$ changes
with $x$, the Fermi surface topology changes.  For $x>0.2$,
Fermi surface crossing occurs on the $(0,0)-(\pi,0)$ line,
giving rise to an electron-like Fermi surface centered at
$(0,0)$ \cite{Ino_ARPES}, as shown in Fig.~\ref{fig1}.  For
$x<0.2$, Fermi-surface crossing occurs on the
$(\pi,0)-(\pi,\pi)$ line, resulting in a hole-like Fermi
surface centered at $(\pi, \pi)$ as in the other cuprates,
as shown in the same figure.  Real Fermi-surface crossing
does not occur at the measurement temperatures ($\sim10$ K)
for most of the samples, however, because a superconducting
gap or a pseudogap is opened on the underlying ``Fermi
surface".  In such a case, the Fermi surface can only be
defined by a minimum gap locus \cite{Ding_FS}.  As for the
dispersive band around $(\pi/2,\pi/2)$, the doping dependent
shift is small.  The doping dependence of the shift is
strong for the band around $(\pi,0)$ and weak for the band
around $(\pi/2,\pi/2)$, as in the case for BSCCO
\cite{Marshall}.

Figure~\ref{fig1} also shows that, for the underdoped
samples, the ``Fermi surface" around $(\pi,0)$ become rather
straight.  This suggests that a one-dimension-like
electronic structure is realized, giving support to the
formation of dynamical stripes.  Very recently, the
evolution of the Fermi surface described above was
demonstrated in terms of two-dimensional intensity plots
\cite{Yoshida}, too, as had been made for LNSCO \cite{Zhou}. 
Here, it should be noted that the spectral weight
distribution around $(\pi,0)$ is rather complicated, making
the determination of the Fermi surface nontrivial.  In
addition to the intrinsic broadness of spectral features and
the opening of the superconducting gap and pseudogap around
$(\pi,0)$, spectral weight remains finite at $E_F$ around
$(\pi,0)$ even for $x=0.3$, where the band is though to be
located well above $E_F$.

\begin{figure}[htb]
\center{\epsfig{file=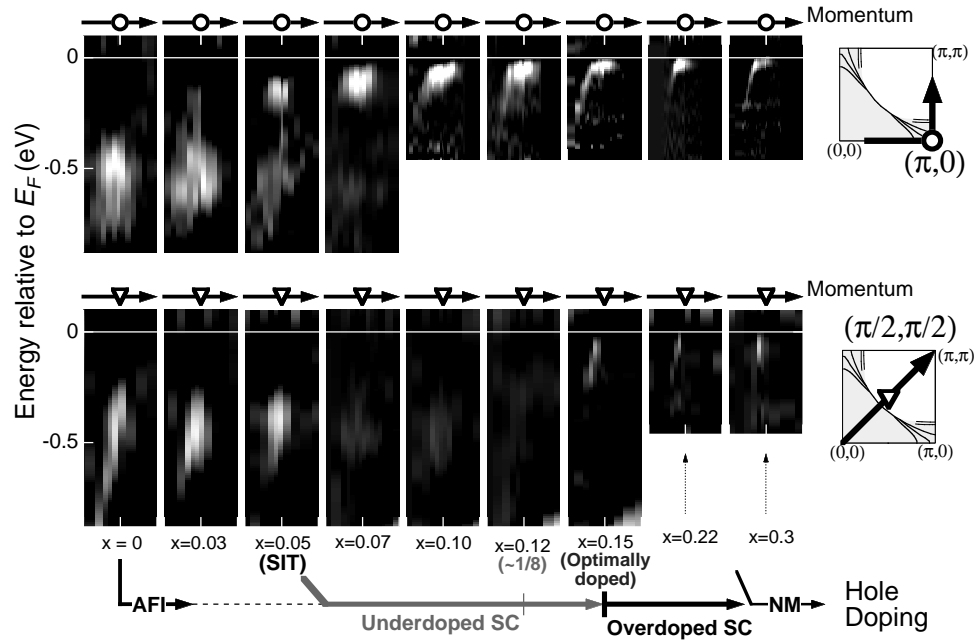,height=85mm}}
\caption{Doping dependence of the band dispersions near $(\pi,0)$ and 
$(\pi/2,\pi/2)$ in LSCO \protect\cite{Ino_ARPES2}.}
\label{fig3}
\end{figure}

The intensity of the dispersive band around $(\pi/2,\pi/2)$
is unusually low in LSCO. In the underdoped samples ($x <
0.15$), most of its spectral weight is transferred to the
higher binding energies of $\sim$0.5 eV, where the
insulating samples ($x\sim0$) show the lower Hubbard band,
as shown in Fig.~\ref{fig3} \cite{Ino_ARPES2,Ino_SIT}. 
Here, the insulating sample $x=0.03$ shows essentially the
same band dispersion as the parent Mott insulator such as
Sr$_2$CuO$_2$Cl$_2$ \cite{Wells}.  The disappearance of the
$(\pi/2,\pi/2)$ band or the ``nodal" quasi-particle (QP)
band can be intuitively understood as due to the presence of
dynamical stripes, which extend along the vertical $(0,\pi)$
or horizontal $(\pi,0)$ direction.  That is, the propagation
of QP along the diagonal direction would be always disturbed
by the hole-poor region of the stripe phase, whereas the
propagation along the Cu-O bond direction would be disturbed
only with the probability of 50 \%.  In contrast to the
nodal QP, the flat band around $(\pi,0)$ is robust against
the decrease of hole concentration.  It persists down to
$x\sim 0.05$, i.e., down to the superconductor-insulator
boundary, and then spectral weight transfer to the lower
Hubbard band at $\sim$0.5 eV below $E_F$ occurs.

\subsection{Chemical potential shift}

According to the stripe picture, the hole density along each
stripe is constant (0.5 per Cu-Cu distance, namely, each
stripe is quarter-filled) whereas the density of the charge
stripes changes with hole doping.  Therefore, as long as the
separation between the stripes is sufficiently large so that
the inter-stripe interaction is negligible, the energy of
the system per hole remains almost constant.  This means
that the chemical potential of electron or hole remains
unchanged with hole doping and hence apparently remains
fixed.  The pinning of the chemical potential has indeed
been observed in LSCO for $x<0.12$ \cite{Ino_XPS}.

Figure~\ref{fig3} demonstrates that ARPES data show
remarkable changes around $x=0.05$ \cite{Ino_SIT}.  The
spectral change at $x\sim0.05$ may reflect the
superconductor-to-insulator transition at this composition
or the change from the dynamical to static stripes because
the coherent spectral weight near $E_F$, which remains
appreciable around $(\pi,0)$, disappears below this $x$. 
This may also reflect the change from the vertical to
diagonal stripes at $x<0.05$ as observed by a recent neutron
scattering study \cite{Yamada2} because the hole motion
along the Cu-O bond direction would be hindered in this
composition range.  In the narrow composition range around
$x=0.05$, the spectrum at every {\bf k} is modeled as a
superposition of that of the insulator ($x\sim0$) and that
of the superconductor ($x\simeq0.07$).

\section{Large pseudogap}

\begin{figure}[htb]
\center{\epsfig{file=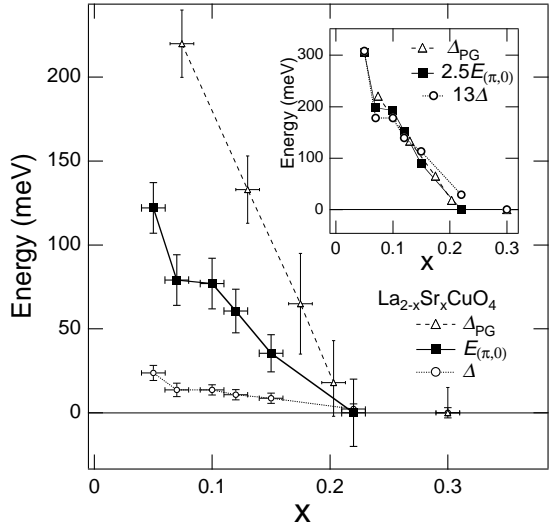,height=70mm}}
\caption{Magnitudes of the superconducting gap $\Delta$,
large pseudogap $\Delta_{PG}$ and the energy position of the
$(\pi,0)$ flat band $E_{(\pi,0)}$ as functions of hole
doping $x$ \protect\cite{Ino_ARPES2}.}
\label{fig4}
\end{figure}

Angle-integrated photoemission (AIPES) studies of LSCO has
revealed a reduction of the density of states (DOS) at $E_F$
and around it below $x\sim 0.2$ \cite{Ino_AIPES}.  This
reduction tracks the decrease of the electronic specific
heat \cite{Loram,Momono} and that of the Pauli component in
the uniform magnetic susceptibility \cite{Nakano} with
decreasing $x$, which occurs in the same composition range. 
These phenomena are consistent with the Fermi-surface
crossing of the flat band at $x\simeq0.2$ since the flat
band is expected to naturally give a high DOS. The downward
shift of the flat band away from $E_F$ with decreasing $x$
reduces the DOS at $E_F$.  If we call the energy region
around $E_F$ with the reduced DOS a ``large pseudogap", the
large pseudogap expands from zero at $x\sim0.2$ to $\sim$0.1
eV at $x\sim0.1$.  As the magnitude $\Delta_{PG}$ of the
large pseudogap increases with decreasing $x$
(Fig.~\ref{fig4}), the temperature $T_{\chi_{\rm max}}$ at
which the magnetic susceptibility takes the maximum
increases \cite{Nakano}, so that the relationship
$\Delta_{PG}/k_BT_{\chi_{\rm max}}\sim3$ holds.

The appearance of a pseudogap below $x=0.2$ and its nearly
linear increase with decreasing hole concentration has been
identified in recent specific heat measurements
\cite{Loram2}.  This observation together with the fact that
$\Delta_{PG}$ is of the order of the in-plane super-exchange
coupling constant $J$ suggest that the antiferromagnetic
coupling between the Cu spins is responsible for the
formation of the large pseudogap.  Such a scenario is
consistent with the $t-J$ model calculation of photoemission
spectra by Jakli{\v c} and Prelov{\v s}ek \cite{Prelovsek}.

\section{Small pseudogap and superconducting gap}

In addition to the large pseudogap, AIPES
\cite{Ino_AIPES,TT} and ARPES \cite{Yoshida} studies have
revealed that a superconducting gap of several meV is opened
at $E_F$.  The gap is identified as a leading edge shift or
a dip in the symmetrized spectra on the Fermi surface,
$A(k_F,\omega)+A(k_F,-\omega)$.  The magnitude of the gap
reaches the maximum near $(\pi,0)$ and satisfies the BCS
relationship of a $d$-wave superconductor
$2\Delta/k_BT_C\sim4$ in the optimally-doped samples whereas
$2\Delta/k_BT_C\gg4$ in the underdoped samples.  Is it found
that $\Delta$ increases with decreasing $x$ and that
$\Delta_{PG}/\Delta$ remains roughly constant in the
underdoped regime (Fig.~\ref{fig4}), implying a correlation
between the large pseudogap and the superconducting gap, and
hence a close connection between the antiferromagnetic
correlations and the superconductivity.  Although our
measurements were made largely in the superconducting state,
we expect that the superconducting gap of the underdoped
samples, which is much larger than $4k_BT_C$, would not
collapse above $T_C$ and remains as a normal-state gap or a
``small pseudogap" as in BSCCO \cite{Loeser_PG}.

\section{Evolution of electronic structure with hole concentration}

The photoemission results described have revealed the
following characteristic hole concentration regions in LSCO:

\noindent (i) $x>0.2$: Normal Fermi-liquid regime with an
electron-like Fermi surface.  The effective mass is enhanced
with decreasing $x$.  The flat band is located above $E_F$. 
The Wilson ratio approaches 2 in the heavily overdoped limit
\cite{Nakano}, indicating that the system is a strongly
correlated two-dimensional Fermi liquid.  The $T_c$ becomes
the highest in this crossover regime.

\noindent (ii) $0.12<x<0.2$: Crossover regime between the
normal Fermi liquid and the dynamical stripe phase.  The
flat band is lowered below $E_F$ and the Fermi surface
becomes hole-like.  The QP is weakened around the nodal
point because of the stripe fluctuations.  The pseudogaps
start to form.

\noindent (iii) $0.05<x<0.12$: Metallic (superconducting)
phase with dynamical stripes.  The Fermi surface becomes
straight lines around $(\pi,0)$, implying one-dimensional
character, and the nodal QP loses most its of spectral
weight.  The chemical potential is pinned associated with
the stripe formation.  The large pseudogaps are fully
developed.  The Wilson ratio becomes $\sim$1 \cite{Nakano},
suggesting that the spectral weight of low-energy
fluctuations are depleted due to the opening of the large
pseudogap.

\noindent
(iv) $x<0.05$: Insulating state with segregated holes. 

\vspace{0.3cm} One may ask the question of how the stripes
and the pseudogaps are related with each other in the
underdoped regime.  Presumably, the hole-poor part of the
stripe phase is more insulator-like and contributes to the
gap-like DOS and hence to the large pseudogap.  The
hole-rich part contributes to the finite DOS at $E_F$ and to
the superconducting gap (and the small pseudogap).

\begin{figure}[htb]
\center{\epsfig{file=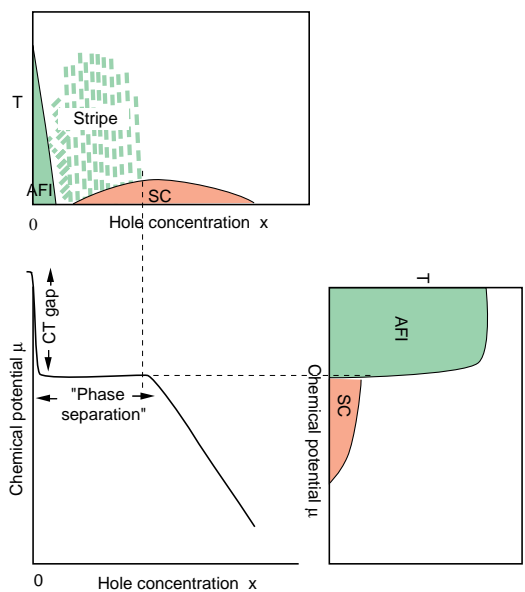,height=80mm}}
\caption{Phase diagram of LSCO plotted against the 
chemical potential $\mu$ 
\protect\cite{Tesanovic}.}
\label{fig5}
\end{figure}

Finally, let us come back to the question of whether the
antiferromagnetic correlation helps the superconductivity or
competes with it.  The similar doping dependence of the
large pseudogap and the superconducting gap may suggest that
the antiferromagnetic correlation, which is likely to be the
origin of the large pseudogap, is responsible for the Cooper
pairing.  On the other hand, if the antiferromagnetic
correlation is the origin of the large pseudogap, it follows
that the antiferromagnetic correlation disrupts the
superconductivity through the reduction of the DOS at $E_F$. 
Then the antiferromagnetic correlation helps and disrupts
the superconductivity simultaneously.  This is analogous to
the situation of phonon-mediated superconductivity in which
too strong electron-phonon interaction causes lattice
instability and thus disrupts the superconductivity.  Now,
it is illuminating to plot the phase diagram of LSCO against
the chemical potential $\mu$ instead of hole concentration
$x$, as shown in Fig.~\ref{fig5}.  Since the chemical
potential does not move for $0<x<0.12$, in the $\mu-T$ phase
diagram, the underdoped region collapses into a phase
boundary line between the superconducting phase and the
antiferromagnetic insulating phase.  As the chemical
potential approaches that of the parent insulator, the $T_c$
monotonously increases due to the increased pairing
potential, but eventually the superconducting state becomes
unstable and is taken over by the antiferromagnetic
insulating phase via a first-order phase transition.  Such
an instability would be inherent in strong coupling
superconductors of magnetic origin \cite{Tesanovic}.  Also,
such a phase diagram implies the degeneracy of the
antiferromagnetic and superconducting phases at the phase
boundary, the situation to which SO(5) theory may be
applicable \cite{SCZhang}.

\begin{acknowledgments}
We would like to thank K. Kishio, T. Kimura and K. Tamasaku
for collaboration in the early stage of this work.  This
work is supported by a Grant-in-Aid for Scientific Research
from the Ministry of Education, Science, Sports and Culture
of Japan, the New Energy and Industrial Technology
Development Organization (NEDO), the U.~S.~DOE's Office of
Basic Energy Science, Division of Material Sciences. 
Experiments were performed at the Stanford Synchrotron
Radiation Laboratory, which is operated by the Office's
Division of Chemical Sciences.
\end{acknowledgments}

\end{document}